\documentclass[11pt,twoside]{article}

\usepackage{asp2006}
\usepackage{epsf}
\usepackage{psfig}
\usepackage{lscape}
\usepackage{graphicx}

\begin{document}
\title{Reconstructing Star Formation Histories of Galaxies}   %%% Fill in title
\author{Uta Fritze$^1$ \& Thomas Lilly$^2$}   %%% Fill in author names
\affil{$^1$ University of Hertfordshire, UK, $^2$ Universit\"at G\"ottingen, Germany}    %%% Fill in author affiliations

\begin{abstract} %%% Abstract to run on from here.
We present a methodological study to find out how far 
back and to what precision star formation histories of 
galaxies can be reconstructed from CMDs, from integrated 
spectra and Lick indices, and from integrated multi-band 
photometry. 
Our evolutionary synthesis models GALEV allow to describe 
the evolution of galaxies in terms of all three approaches 
and we have assumed typical observational uncertainties for 
each of them and then investigated to what extent and 
accuracy different star formation histories can be 
discriminated. For a field in the LMC bar region with both 
a deep CMD from HST observations and a trailing slit spectrum 
across exactly the same field of view we could test our 
modelling results against real data. 
\end{abstract}

%%% MAIN BODY OF TEXT GOES HERE. CONSULT "INSTRUCTIONS FOR AUTHORS USING
%%% LATEX2E MARKUP", SECTIONS 2.3-2.6 FOR HELP WITH EQUATIONS, FIGURES,
%%% AND TABLES.

\section{GALEV Evolutionary Synthesis Models}   
GALEV evolutionary synthesis models describe the evolution of stellar populations from the onset of star formation ({\bf SF}) 
over a Hubble time. They describe single burst simple stellar populations ({\bf SSPs}) like star clusters, which consist of only one generation of stars, as well as galaxies, which are composite stellar populations, both in terms of stellar ages and in terms of stellar metallicities. The input physics for stars and gas they use comprise the latest stellar evolutionary tracks/isochrones from the Padova group, stellar model atmospheres from Kurucz/Lejeune, Lick stellar absorption indices, gaseous emission in terms of lines and continuum, stellar yields for PNe, SNII, SNIa (carbon deflagration white dwarf binaries) for elements from H, He, C, N, O, Mg, $\dots$, through Fe for 5 metallicities in the range ${\rm -1.7 \leq [Fe/H] \leq +0.4}$ and solar scaled abundances (cf. Anders \& Fritze 2003, Lilly \& Fritze 200xxx, Lindner, Fritze, Fricke 1999 for details). {\em  All these pieces of input physics depend significantly on metallicity, and, hence, does the output physics}. 

The output physics of GALEV models comprises the time evolution of color magnitude diagrams ({\bf CMD}s) for star clusters and galaxies and the time evolution of spectra (90 \AA $\dots$ 160 $\mu$m), emission line strengths, luminosities (U $\dots$ K) in a large variety of filter systems (Johnson, HST, Washington, Stroemgren, $\dots$), colors, absorption features (Mg$_2$, Mgb, Fe5270, Fe5335, TiO1, TiO2, $\dots$), stellar and gaseous masses, mass-to-light ratios, ISM abundances (H, He, C, N, O, Mg, Ne, $\dots$, Fe) and abundance ratios ([C/O], [N/O], [Mg/Fe], $\dots$).

The metallicity dependence of the output physics is best seen on SSP models, which clearly show how the evolution in various colors depends on metallicity in a way very specific for each color, how the fading in various bands and the Lick index evolution, {\em even for the Balmer indices}, depend on metallicity (cf. Schulz, Fritze, Fricke 2002, Anders \& Fritze 2003, Lilly \& Fritze 2006). GALEV models for SSPs of various metallicities can be retrieved from {\em http://star-www.herts.ac.uk/}$^{\sim}${\em ufritze/UH}$_{-}${\em data}.

Like for any other evolutionary synthesis code (Bruzual/Charlot, Pegase, Starburst99), the star formation history ({\bf SFH}) is {\em the} basic parameter for GALEV models for galaxies. Simplified parametrisations are used for the various spectral types of galaxies, similar to the ones suggested by Sandage (1986): exponentially declining SF rates with an e-folding timescale of 1 Gyr for the classical elliptical model, and SF rates tied to the evolving gas content with efficiencies declining from S0 through Sc galaxies, and, finally, constant SF rates for pure disk/Sd galaxies. {\em These simplified SFHs are tightly constrained by the required agreement of spectral and chemical properties of local galaxy samples/templates of the respective type}. 

A key feature of GALEV evolutionary synthesis models is that they {\em simultaneously treat the chemical and spectral evolution} of galaxies, which has two advantages: first, a larger number of predicted observables (ISM abundances and spectral features of the stellar population) with the same small number of free parameters (SFH and stellar IMF) as models treating only one aspect, and ${\rm 2^{nd}}$, they allow for what we call the {\em chemically consistent} approach. This means that {\em we monitor the ISM abundance at the birth of each stellar generation and describe each stellar generation with input physics appropriate for its initial metallicity}. We thus consistently account for the increasing initial abundances of successive stellar generations, can follow the successive build-up of the composite stellar metallicity distribution in galaxies and account for the locally observed stellar metallicity range in galaxies. In particular, {\em chemically consistent models account for the increasing importance of low metallicity stellar (sub-)populations in local late-type and dwarf galaxies and in the less evolved galaxies observed at high redshift}. 

We note that stellar yields at low metallicity significantly differ from solar metallicity yields, and that stellar yield ratios [C/O], [N/O], [Mg/Fe], $\dots$ change with metallicity. Elements with different nucleosynthetic origin are restored to the ISM on different timescales with the timescale being shorter for SNII/$\alpha$-elements than for intermediate mass elements like C or N, and longest for the SNIa element Fe. This causes {\em all ISM abundance ratios to depend on the SFH of the galaxy. Via the SFH, galaxy evolution and stellar evolution get intimately coupled} and, in principle, stellar evolutionary tracks/isochrones, yields, and model atmospheres are required for the full range of element ratios. Since these are not available at present, we restrict our models to solar scaled abundance ratios. 

\section{CMDs versus Integrated Photometry versus Spectroscopy}
For single burst SSPs and for galaxies, GALEV models calculate the time evolution of the stellar population in the HR diagram in the first place. These are easily transformed into CMDs in arbitrary passband combinations. These model CMDs in various passbands can be used to optimise observational strategies, e.g. when aiming at disentangling age, metallicity and extinction degeneracies for star clusters or revealing different age-metallicity subpopulations in galaxies. {\em A long wavelength basis is essential in any case.}

Fig. 1 shows a comparison of CMDs in various passband combinations for two SSPs of age-metallicity combinations (500 Myr, Z$=0.008$) and (600 Myr, Z$=0.004$). These combinations are indistinguishable both in V$-$I and V$-$K but separate clearly in U$-$V. Extensive tests have shown that {\em for young stellar populations the U-band is important for age determinations and for old ($>3$ Gyr) populations the K-band is important for metallicity determinations.} 

\begin{figure}[!h]
\begin{center}
\includegraphics[width=4.cm]{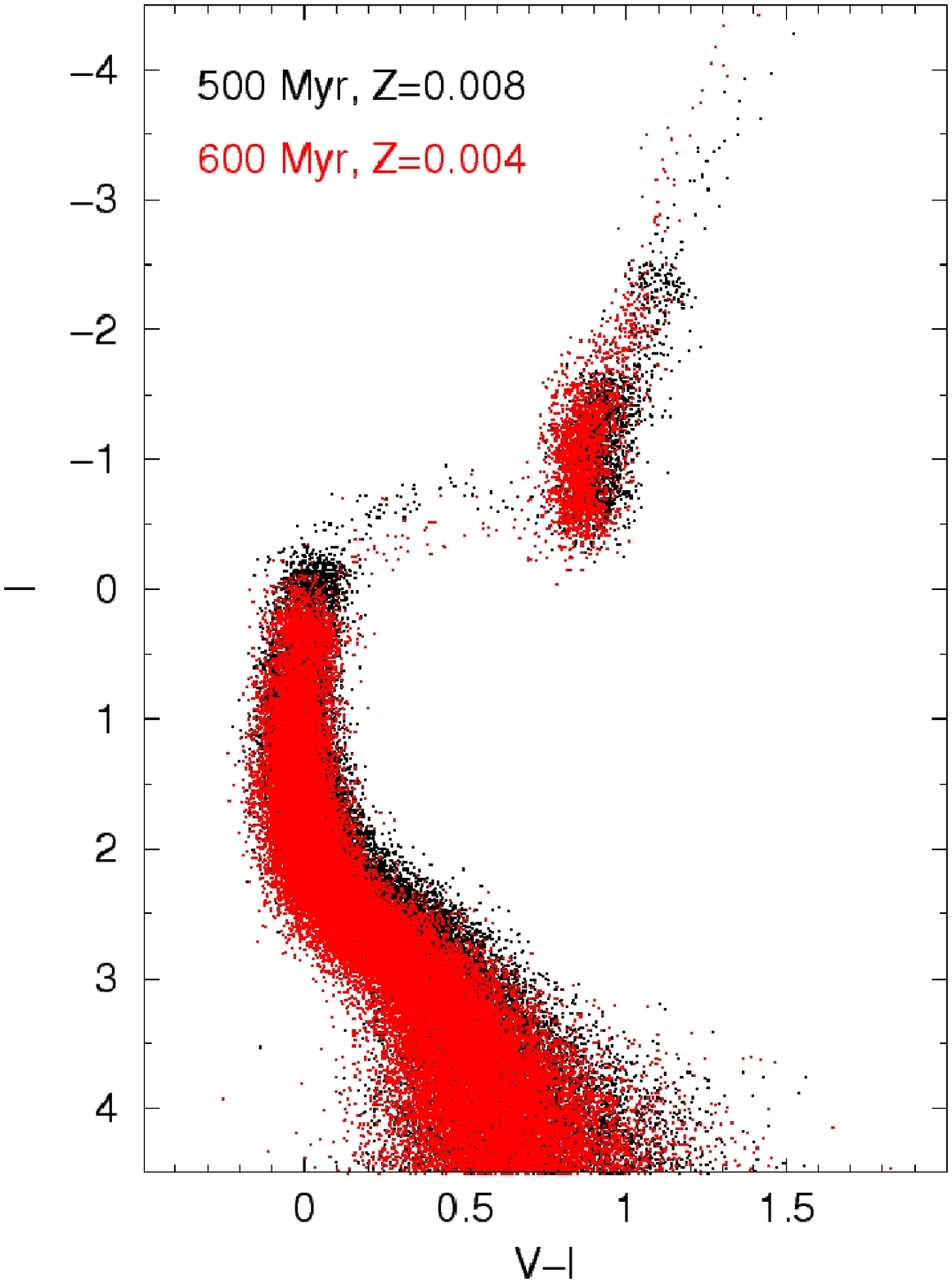}
\includegraphics[width=4.cm]{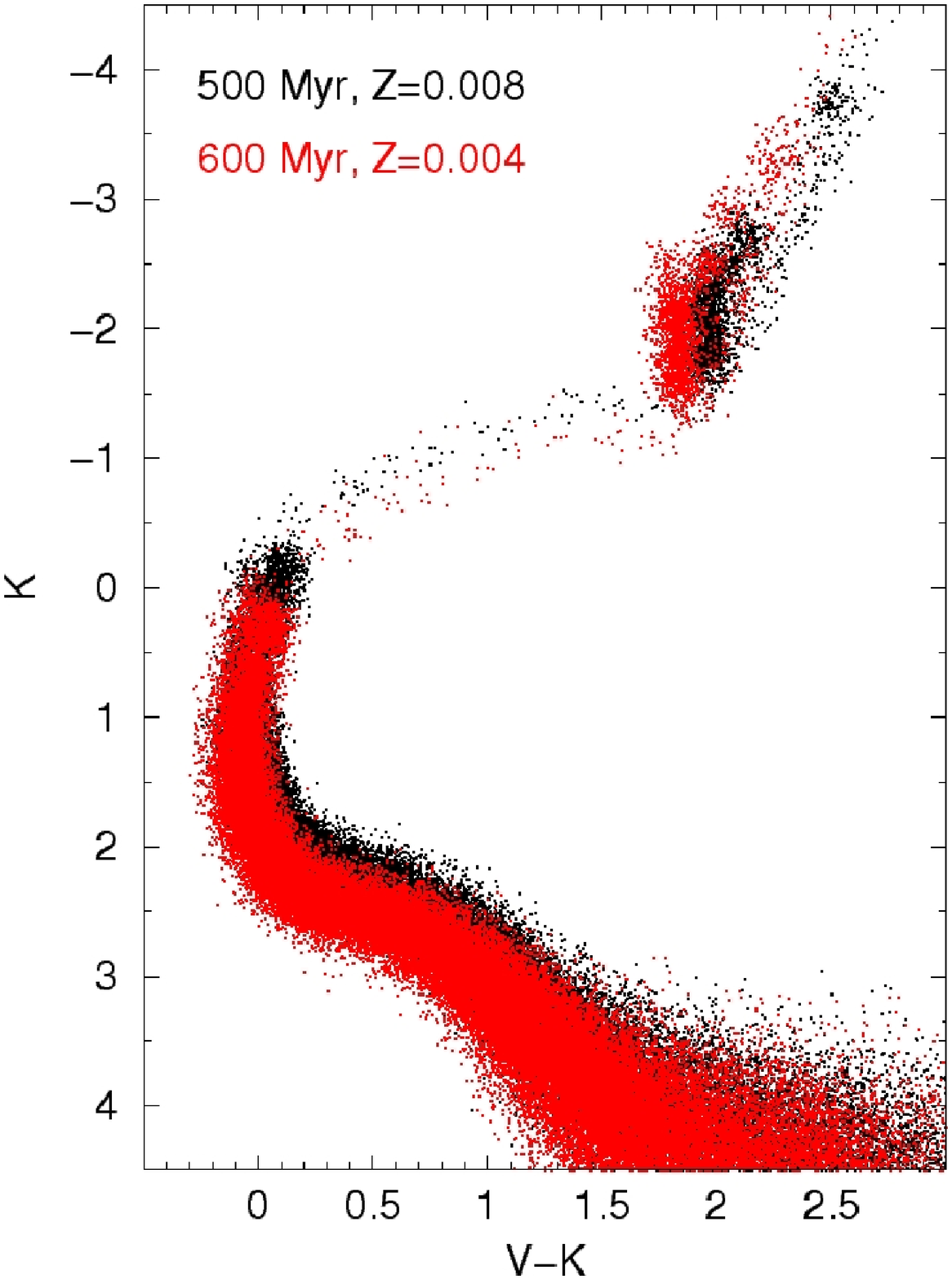}
\includegraphics[width=4.cm]{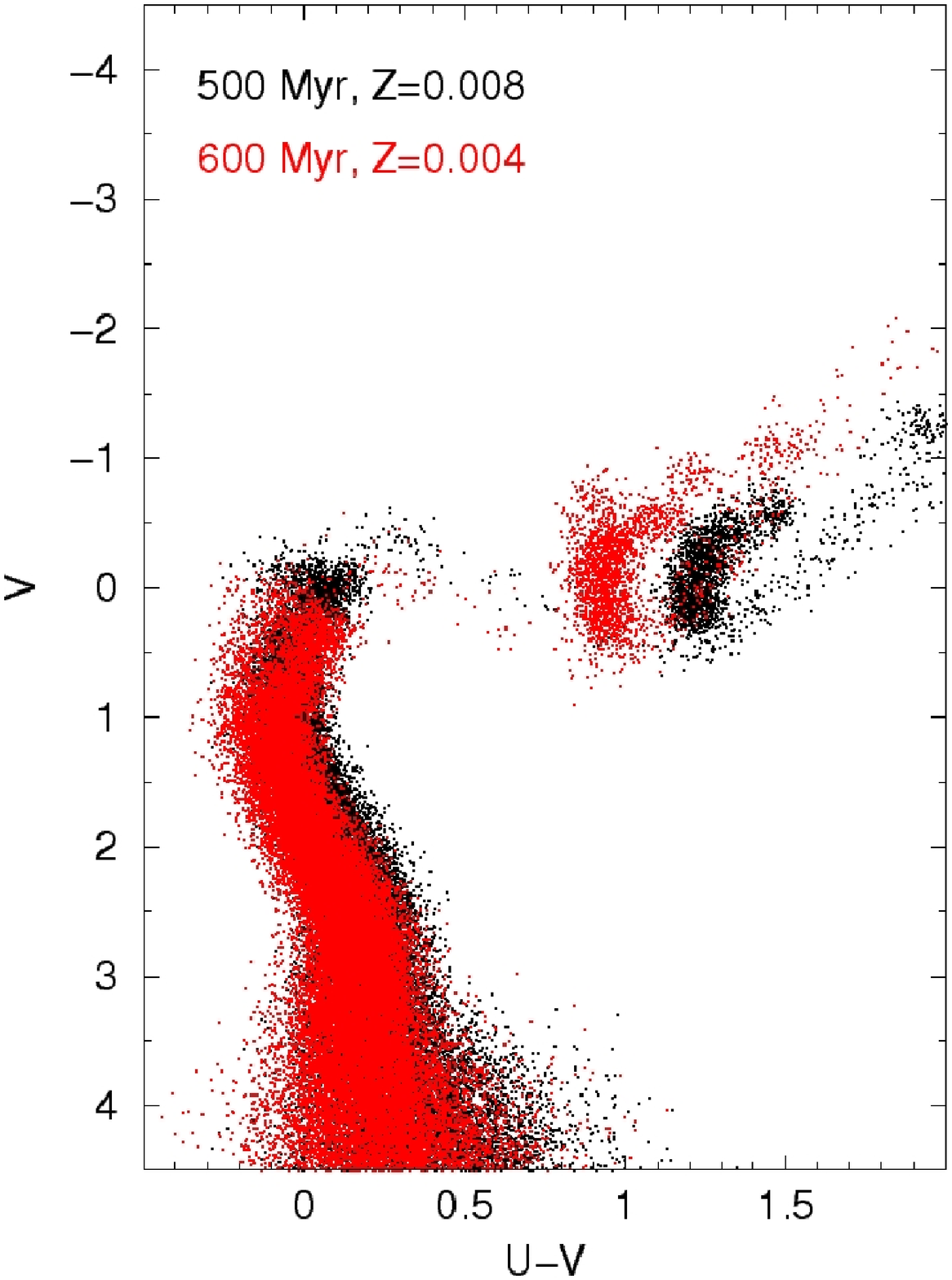}
\end{center}
\caption{Three CMDs in V$-$I, V$-$K, and U$-$V, each showing two SSPs of age-metallicity combinations (500 Myr, Z$=0.008$) and (600 Myr, Z$=0.004$). Both SSPs cannot be distinguished in V$-$I and V$-$K, but clearly split up in U$-$V.}
\end{figure}

Since GALEV models calculate the time evolution of CMDs as well as of integrated spectral and photometric properties, they allow for a methodological investigation of the question how far back and to what accuracy all three methods allow to trace back the SFHs of galaxies. We have done this in a theoretical study accounting for typical observational uncertainties in the three approaches (e.g. 0.05 mag in colors, 0.1 \AA in Lick indices, and S/N=5 $-$ 10 in spectra) using simplified SFHs and fixed metallicity for clarity (cf. Lilly \& Fritze 2005a, b), and in an observational study, the LMC bar project. The LMC bar project is a collaboration with C. Gallart, D. Alloin, P. Demarque, and others and compares the SFHs derived by various groups from a deep HST CMD of a field in the LMC bar region (known as the Coimbra experiment, cf. Holtzman 02) and the SFHs derived from a VLT spectrum obtained by trailing a slit across exactly the same field. 

Results from our theoretical studies are presented in Lilly \& Fritze (2005a) and summarised as follows: \\
\noindent
On the basis of CMDs, two simple SF scenarios, one with a constant SF over a period of 2 Gyr, the other forming the same amount of stars in two short bursts at the beginning and end of this period, can at best be discriminated for $\leq 4$ Gyr after the end of SF, and only if the SFR really is zero after the first 2 Gyr.\\
\noindent On the basis of colors, different SF scenarios can again at best be discriminated for 1 $-$ 4 Gyr after the end of SF and there is absolutely no way to reveal the SFH prior to a burst or the last major epoch of SF.\\
\noindent On the basis of Lick indices, different SF scenarios can again be discriminated for maximally 1 $-$ 4 Gyr after the end of SF (using H$_{\beta}$, e.g.).\\
\noindent On the basis of low/intermediate resolution spectroscopy, there is again, no way to reveal the SFH prior to a burst or the last major epoch of SF and different scenarios cannot be discriminated any more after 1 $-$ 4 Gyr. High resolution high S/N spectroscopy and narrow band indices will do somewhat better with e.g. MgII2798 and CaIIK3933 being visible for $\sim 2$ and $2 - 6$ Gyr, respectively (see also G. Bruzual, this volume).

\noindent
The longer lookback times in all cases refer to a long wavelength basis and particularly ideal SF scenarios, all estimates are at fixed metallicity. In the presence of metallicity differences between the two scenarios or metallicity evolution between the SF episodes within a given scenario, the uncertainty increases and can only partly be overcome with the use of a very long wavelength basis (UV through NIR).

\noindent
Hence, we conclude that {\em SFH details can be recovered with similar accuracy from broad band colors or SEDs, from Lick indices, from low/intermediate resolution spectroscopy} (see also Cardiel et al. 2003){\em , and from CMDs} (see also E. Tolstoy, this volume).

\noindent
The observational study comparing SFHs for an LMC bar field derived from a deep HST CMD by Smecker-Hane et al. (2002) and from our VLT spectrum (4000 $-$ 8000 \AA) yielded the following results (see Lilly \& Fritze 2005b for details): \\
\noindent The CMD that we modeled on the basis of Smecker-Hane et al.'s fairly complex SFH agrees very well with the one observed with HST. A quantitative assessment of this comparison still needs to be done.\\
\noindent We constructed a toy model having only three different phases of constant SFRs: the first phase over a period of 10 Gyr, the second one over the next 3 Gyr and the third one over the last Gyr. Comparing model spectra with the VLT spectrum, we found that only relative amounts of stars formed in the three phases are relevant for the degree of agreement between model and observed spectra. How the SFR is distributed within each phase does not matter at all -- whether it is concentrated at the onset or end of a phase or constant over all of the respective phase.\\
\noindent The agreement with the observed spectrum that we reached with an appropriate three phase toy model is comparable to the agreement reached with Smecker-Hane et al.'s more complicated SFH. 

\noindent
{\em We conclude that 3 (or eventually 4) phases of SF can be discriminated by either method, CMD analysis or spectral analysis, not more, and it is only the relative amounts of SF in each of the three phases that determines the final CMD or spectrum, not the distribution of the SFR within each of the three phases.} 

From all of these studies, we finally conclude: {\em both from CMDs and integrated light (multi-band photometry as well as spectroscopy)\\
$\bullet$ SFRs during the last Gyr are very precisely recovered,\\
$\bullet$ SFRs between 1 and 3 Gyr ago are roughly recovered, and \\
$\bullet$ SFRs longer than 3 to 5 Gyr ago are only vaguely recovered,}\\ 
with ''roughly'' and ''vaguely'' meaning that only the relative amounts of stars formed in the three different intervals can be determined, but not the details of the distribution of this SF within the respective interval. {\em Essential for the accuracy of the derived SFHs is a long wavelength basis. Photometry over a long wavelength range (ideally UV or U-band through NIR) very well compensates for the higher amount of information contained in a spectrum, which usually covers a much smaller wavelength range.}  

\section{Star Clusters and Globular Clusters}
In recent years, a very interesting complementary approach has come up: to assess galaxy SFHs over cosmological timescales through studies of their star cluster and globular cluster ({\bf GC}) systems (cf. Brodie \& Strader 2006). It has become clear that star cluster formation is an important mode of SF in general, and the dominant mode in starbursts. In the Tadpole and Mice interacting starburst galaxies, as much as 70 \% of the blue light is contributed by star clusters as opposed to field stars, implying that $>35$ \% of SF is star cluster formation (de Grijs et al. 2003). The formation of massive and long-lived clusters only occurs in periods or regions with high SF efficiencies, as e.g. evidenced by the famous age gap in the LMC star cluster population, where SF and chemical enrichment continued on a low level, while star cluster formation ceased for $\sim 7$ Gyr. Hydrodynamical models show that the formation of GCs, i.e. star clusters massive and strongly bound enough to be able to survive as bound entities over $\geq 10$ Gyr, requires SF efficiencies $\geq 30$ \%, with SF efficiency in this context defined as the mass ratio of stars formed from a given reservoir of gas (Brown,  Burkert \& Truran 1995, Li et al. 2004). The highest SF efficiencies are found in massive gas-rich mergers, many of which are luminous or ultraluminous IR galaxies, and believed to result from the high ambient pressure in these systems (Jog \& Das 1992, 1996). 

Star clusters are simple stellar populations, easy to model and easy to analyse on a one-by-one basis after careful background subtraction. Via multi-band photometry they are accessible out to Virgo cluster distances and beyond. 
The ages and metallicities of young star cluster populations are prime tracers of recent/ongoing SF in galaxies. Ages and metallicities of globular cluster populations are tracers of the violent phases of SF over a Hubble time. In both cases, the star clusters are much better suited to trace back the SFHs of their parent galaxies than the integrated light, which inevitably is dominated by the last major epoch of SF, as we have shown before. 

The key is to obtain ages and metallicities from multi-band spectral energy distributions (UV/U, B, V, R, I, J, H, K). Colors can be transformed to metallicities only if a fixed age is assumed and to ages only when a fixed metallicity is assumed. Analyses of a full SED allows to obtain age and metallicity (and internal extinction)  of a star cluster independently. Multi-band imaging in combination with appropriate SED analysis methods then yields independent ages and metallicities of wholesale star cluster or GC populations and allows to construct their age and metallicity distributions, which, in turn, give valuable information about the (violent) SFH of the parent galaxy. In Anders et al. 2004a, b, we have developed an SED Analysis Tool that, on the basis of a large number of GALEV SSP models with a wide range in metallicities (${\rm -1.7 \leq [Fe/H] \leq +0.4}$), ages (4 Myr $\dots$ 14 Gyr), and internal extinctions (${\rm 0 \leq E(B-V) \leq 1}$, assuming the starburst extinction law from Calzetti et al. (2000), and a $\chi^2$-method derives the ages, metallicities, extinction values, and masses of star clusters. It not only looks for {\em the} best fit model, but evaluates the quality of all fits to obtain $1 \sigma$ uncertainty ranges for all the above parameters. Extensive artificial star cluster analysis has shown that 4 passbands are required for young star cluster populations, for which internal dust usually plays an important role, with U or B, V, R or I, and a NIR band being ideal combinations. In intermediate age and old GC systems dust usually does not play a role any more. Here three passbands, again covering U/B through NIR, are enough. Our method meanwhile has been successfully applied by us and other groups to numerous star cluster systems of all ages (Anders et al. 2004b, de Grijs et al. 2003, 2005, Lamers et al. 2006).

The ACS Virgo Cluster Project has obtained B$-$I color distributions for the GC systems in all Virgo E/S0s (Peng et al. 2006). Many E/S0s show bimodal B$-$I color distributions with a fairly universal blue peak unisonously attributed to an old and metal-poor GC population and a red peak, variable in color and relative height from galaxy to galaxy, and with some controversies remaining on the age and metallicity of its cluster population due to the well-known age-metallicty degeneracy of optical colors. GALEV models clearly show that, indeed, the typical color of a red peak (V$-$I=1.2) can be due to clusters with age and metallicity combinations as different as (2 Gyr, [Fe/H]$=+0.4$) and (14 Gyr, [Fe/H]$=-1.7$). These two populations will, however, split up in V$-$K by as much as ${\rm V-K =3.5}$ in the first case and ${\rm V-K =2.3}$ in the second. With additional K-band imaging, GC ages and metallicities can be derived to accuracies of ${\rm \Delta age/age \sim 0.2}$ and 0.2 dex, respectively.

{\em To conclude: Age and metallicity distributions of star and globular cluster systems are prime upcoming tracers of their parent galaxies' star formation, chemical enrichment, and mass assembly histories over the full cosmological timespan.}

%\acknowledgements %%% Text of acknowledgements runs on after this command.

%%% THE BIBLIOGRAPHY
%%%
%%% CONSULT SECTION 3 OF "INSTRUCTIONS FOR AUTHORS" FOR HOW TO USE NATBIB.
%%% AUTHORS ARE ENCOURAGED TO USE EITHER THE "THEBIBLIOGRAPY" ENVIRONMENT
%%% BY UNCOMMENTING (DELETING THE "%" SYMBOL) THE COMMANDS BELOW, OR BY
%%% USING THE BIBTEX ENVIRONMENT. TO FIND OUT WHICH IS APPLICABLE TO YOUR
%%% CONTRIBUTION, CONSULT THE VOLUME EDITORS FOR YOUR PROCEEDINGS.
%%%

\end{document}